\documentstyle[prl,multicol,aps,epsfig]{revtex}
\begin{document}
\draft
\title
{Bound States in the
One-dimensional Hubbard Model}
\author{Daniel Braak$^{1,2}$ and Natan Andrei$^{1}$}
\address{$^1$ Department of Physics and Astronomy, Rutgers University,
Piscataway, NJ 08855}
\address{$^2$ NEC Research Institute, 4 Independence Way, Princeton,
NJ 08540}
\date{\today}
\maketitle
\begin{abstract}
The Bethe Ansatz equations for the one-dimensional Hubbard model
are reexamined. A new procedure is introduced to properly
include bound states. 
The corrected equations
lead to  new elementary excitations away from half-filling. 
\end{abstract}
\def\p{\phi}
\def\P{\Phi}
\def\e{\eta}
\def\ep{\epsilon}
\def\vep{\varepsilon}
\def\ps{\psi}
\def\a{\alpha}
\def\ab{{\tilde{a}}}
\def\b{\beta}
\def\y{{\tilde{y}}}
\def\k{\kappa}
\def\psd{\ps^{\dagger}}
\def\psdt{{\tilde{\ps}}^{\dagger}}
\def\pst{\tilde{\ps}}
\def\di{d^{\dagger}}
\def\d{\delta}
\def\g{\gamma}
\def\be{\begin{equation}}
\def\ee{\end{equation}}
\def\bdm{\begin{displaymath}}
\def\edm{\end{displaymath}}
\def\s{\sigma}
\def\bea{\begin{eqnarray}}
\def\eea{\end{eqnarray}}
\def\bear{\begin{array}}
\def\eear{\end{array}}
\def\nn{\nonumber}
\def\x{\chi}
\def\ra{\rightarrow}
\def\r{\rho}
\def\s{\sigma}
\def\vs{\vec{\s}}
\def\ss{\vec{\sigma}_j\cdot \vec{S}}
\def\S{\bar{S}}
\def\Sr{\tilde{S}}
\def\z{\zeta}
\def\Z{\bar{Z}}
\def\Zr{\tilde{Z}}
\def\ta{\tilde{A}}
\def\t{\tau}
\def\th{\theta}
\def\Th{\Theta}
\def\l{\lambda}
\def\L{\Lambda}
\def\K{{\cal K}}
\def\Li{{\cal L}}
\def\E{{\cal E}}
\def\ki{\frac{k_i}{\L}}
\def\kj{\frac{k_j}{\L}}
\def\ua{\uparrow}
\def\da{\downarrow}
\def\rd{\textrm{d}}
\def\id{{\mathbf{1}}}
\def\ph{\phantom}
\def\O{{\cal{O}}}
\def\o{\omega}
\def\H{\tilde{H}}
\def\arsinh{{\textrm{arsinh}}}
\def\cotanh{{\textrm{cotanh}}}
\def\arcsin{{\textrm{arcsin}}}
\def\sech{{\textrm{sech}}}
\def\vu{\frac{4}{u}}
\def\uv{\frac{u}{4}}
\def\uz{\frac{u}{2}}
\def\cz{\frac{c}{2}}
\def\dx{\Delta\x}
\def\ecl{e^{-\frac{|c|}{2}L}}
\def\eclp{e^{\frac{|c|}{2}L}}
\def\eul{e^{-\frac{|u|}{4}L}}
\def\eulp{e^{\frac{|u|}{4}L}}
\def\ekl{e^{-\xi L}}
\def\eklp{e^{\xi L}}
\def\ekld{e^{-2\xi L}}
\def\ln{{\textrm{ln}}}
\def\G{{\hat{G}}}
\def\ran{\rangle}
\begin{multicols}{2}
The one-dimensional Hubbard 
hamiltonian,
\be
H=\sum_{i=1}^L-t\left[\psd_{\s,i+1}\ps_{\s,i} +
h.c.\right]+Un_{\ua,i}n_{\da,i}
\label{sham}
\ee
describes fermions moving on a ring of length $L$ with
 an on-site interaction $U$.
The model was solved by Lieb and Wu \cite{liebwu} using  Yang's
 generalized
Bethe Ansatz  method (BA) \cite{yang}. They derived
the following
equations,
 describing an eigenstate of (\ref{sham}) in the sector
with $N$ particles and total spin $S=\frac{1}{2}(N-2M)$,
\be
e^{ik_jL}=\prod_{\d=1}^M\frac{\l_\d-\sin k_j-i\frac{u}{4}}
{\l_\d-\sin k_j+i\frac{u}{4}}
\label{lw1}
\ee 
\be
\prod_{\d\neq\g}^{M}\frac{\l_\g-\l_\d-i\frac{u}{2}}
{\l_\g-\l_\d+i\frac{u}{2}}
=\prod_{j=1}^{N}\frac{\l_\g-\sin k_j-i\frac{u}{4}}
{\l_\g-\sin k_j+i\frac{u}{4}}
\label{lw2}
\ee
where  $u=U/t$. 
The $M$ ``spin momenta'', $\l_\g$, and the $N$ "charge momenta",  $k_j$,
describe
the many-body correlations of this
state. Once these parameters are determined from the BA equations the energy
 and momentum of the state are,
$E=-2t\sum_j \cos k_j $, $ P=\sum_j k_j.$
See ref \onlinecite{and} for a detailed derivation. 

We shall argue that the equations as they stand 
are incomplete.
The problem arises when one considers complex momenta solutions
which describe bound states and double occupancy. 
In the standard approach \cite{taka,woyn} these momenta, 
$k^{\pm}=q\pm
i\xi$, are driven, in the 
infinite volume limit, $L\ra\infty$, to 
``string positions''
\be
\sin k_l^\pm=\L_l\mp i\uv  + \vep,
\label{stri}
\ee
where $\vep=\O(\eul)$ and
 $\L_l$ is identified with an element in the
set of spin momenta $\{\l_\g\}$.
This expression is then inserted  into (\ref{lw1},\ref{lw2}).
Simple manipulations lead to,
\bea
e^{ik_jL}=\prod_{\d=1}^{M^u}\frac{\l_\g-\sin k_j-i\frac{u}{4}}
{\l_\g-\sin k_j+i\frac{u}{4}}
\prod_{l=1}^{N^b}\frac{\L_l -\sin k_j -i\frac{u}{4}}
{\L_l -\sin k_j +i\frac{u}{4}}(1+\vep)
\label{lw1rn}\\
\prod_{\d\neq\g}^{M^u}\frac{\l_\g-\l_\d-i\frac{u}{2}}
{\l_\g-\l_\d+i\frac{u}{2}}
=\prod_{j=1}^{N^u}\frac{\l_\g-\sin k_j-i\frac{u}{4}}
{\l_\g-\sin k_j+i\frac{u}{4}}(1+\vep)
\label{lw2rn}\\
\prod_{\d=1}^{M^u}\frac{\L_l-\l_\d-i\frac{u}{2}}
{\L_l-\l_\d+i\frac{u}{2}}
=\prod_{j=1}^{N^u}\frac{\L_l-\sin k_j-i\frac{u}{4}}
{\L_l-\sin k_j+i\frac{u}{4}}(e^{i\varphi_l}+\vep)
\label{bslan}\\
e^{2iq_lL}=\prod_{\g=1}^{M^u}\frac{\l_\g-\L_l-i\frac{u}{2}}{
\l_\g-\L_l+i\frac{u}{2}}
\prod_{n\neq l}^{N^b}\frac{\L_n-\L_l-i\frac{u}{2}}{ 
\L_n-\L_l+i\frac{u}{2}}(e^{i\varphi_l}+\vep)
\label{lw1arn}
\eea
The $k_j$, $j=1\ldots N^u$ denote the real momenta, $k^{\pm}_l=q_l\pm
i\xi_l$, $l=1\ldots N^b$, are paired with the $\L_l$ and 
$\l_\g$, $\g=1\ldots M^u$, runs over the set of unpaired
spin-parameters, $N=N^u+2N^b$ and $M=M^u+N^b$.
The phase $e^{i\varphi_l}$ is defined as
\[
e^{i\varphi_l}=-\frac{\L_l-\sin k_l^+ -i\uv}
{\L_l-\sin k_l^- +i\uv}.
\]
We show now that the "string Lieb-Wu" equations,
(\ref{lw1rn},\ref{lw2rn},\ref{bslan},\ref{lw1arn}),
 cannot describe {\it all}
(highest weight) states in the infinite volume limit. In other words,
the 
infinite volume
limit cannot
 be obtained as  $L\ra\infty$ limit of the solutions to
(\ref{lw1},\ref{lw2}) for finite $L$. The reason is, that 
equations (\ref{bslan}),  which do not contain $L$ explicitly,
become redundant for  $L\ra\infty$. Maintaining them, as is commonly
done, 
cuts out 
a set of eigenstates which {\it do} exist in the infinite system.
To show this, we use the algebraic Bethe ansatz;  assume a pair of  complex
 conjugated momenta $k^+, k^-$ related to the
spin momentum $\L$ by (\ref{stri}).
An eigenstate  of the corresponding inhomogeneous
transfer matrix $Z(\mu)$ with (arbitrary) spectral parameter $\mu$ 
 then has the form (see e.g. \cite{and}): $
|\ps\ran = \prod_{\g=1}^{M^u}B(\l_\g)B(\L)|\o\ran$
for  total spin $S=\frac{1}{2}(N^u-2M^u)$, 
where the $M^u$ creation operators $B(\l_\g)$, acting on the
ferromagnetic vacuum $|\o\ran$, create $M^u$ $\da$-spins, and
$B(\L)$ creates the $\da$-spin of the bound state. 
By explicit calculation we find that $B(\L)$ acting on $|\o\ran$
 diverges as $\eulp$
for $L\ra\infty$, whereas the $B$'s not associated with the
complex pair do not diverge. Normalizing $|\ps\ran$ by
multiplication with $\eul$ yields
accordingly an  exponential suppression of vectors of the form
$\prod_{\d}B(\l_\d)|\o\ran$ which {\it do not} contain
$B(\L)$ among the $B(\l_\d)$. Now the equation (\ref{bslan}) for $\L$
is necessary to cancel an ``unwanted term'' in the eigenvalue
equation for 
$|\ps\ran$:
\bea
&&(Z(\mu)-E(\mu))|\ps\ran= \nn\\ 
&&\sum_{\g}^{M^u}\a_\g\prod^{M^u}_{\d\neq\g}B(\l_\d)B(\L)B(\mu)|\o\ran
+\a_0\prod_{\g}^{M^u}B(\l_\g)B(\mu)|\o\ran. \nn
\eea
It ensures, in particular, that $\a_0=0$. But, as argued previously,
the vector $\prod_{\g} B(\l_\g)B(\mu)|\o\ran$ is
projected out in the infinite volume limit (exponentially suppressed
for
finite $L$), and therefore (\ref{bslan}) is not necessary for $|\ps\ran$
to be an eigenvector of $Z(\mu)$ for $L\ra\infty$. That means that only a 
subset of all states allowed in the infinite volume can
be  generated by the $L\ra\infty$ limit of solutions to 
(\ref{lw1},\ref{lw2}) for
finite $L$.

We proceed to introduce a procedure that generates all (highest weight)
eigenstates of 
the hamiltonian in the infinite volume limit. We begin by 
 deriving the
elementary bound states  of the hamiltonian,
 and  constructing the S-matrices 
between them and the unbound particles. The S-matrices will be employed
to derive the consistent BA equations.
The  normalizable
two-particle bound state solutions of the hamiltonian
have the form,
\be
F^{b}(n_1,n_2)=A_{a_1a_2}e^{iq(n_1+n_2)}e^{-\xi(q)|n_1-n_2|}
\label{bst}
\ee 
where  the parameters  $q$ and $\xi$ are related by, $
\sinh\xi(q)=-\frac{u}{4\cos q}$.
 As  $\xi\ge 0$,
the range for $q$ in the interval $[-\pi,\pi]$ is: $|q|>\frac{\pi}{2}$ for
$U>0$ and $|q|<\frac{\pi}{2}$ for $U<0$. In the former case the real parameter
 $q$ (which is half
the quantum mechanical momentum: $p=2q$) lies near the center of the
Brillouin zone and in the latter near the edges.

The bound states are necessarily  spin singlets, and their
 energy  is $E=-4t\cos q\cosh\xi=U\coth\xi(q)$.
Thus, it depends directly on
$U$, and is negative in the attractive case ($U<0$) and positive in the
repulsive case.  Moreover there is a
gap
in the spectrum: $E(q)\ge |U|$ for $U>0$, and $E(q)\le -|U|$ for $U<0$.
The limiting values $\pm |U|$ correspond to $q\ra \pm(\pi/2)$, with
 $\xi(q)$ tending to
infinity at these points.

Typically, 2-particle eigenstates of the model have the form (denote $n_{1,2}=
n_1-n_2$ and ${\cal A}$ the antisymmetrizer),
\bea
&&F(n_1,n_2)= \nn \\
 &&={\cal A}e^{ik_1 n_1}e^{ik_2n_2}[A_{a_1a_2}\th(n_{2,1})
+S_{12}A_{a_1a_2}\th(n_{1,2})].
\label{bst2}
\eea
with $S_{12} \equiv S^{uu}_{12}$ being the S-matrix between {\it
unbound} 
particle 1 and 2
carrying real momenta $k_1$ and $k_2$. It
reads,
\bea
S^{uu}_{12}(k_1,k_2)=\frac{\sin k_1-\sin k_2 +i\frac{u}{2}P_{12}}{
\sin k_1-\sin k_2 + i\frac{u}{2}} 
\label{free}
\eea
We can  view the bound state as corresponding to 
complex momenta $k_1=k^-=q-i\xi$, $k_2=k^+=q+i\xi$, with 
$\sin k^{\pm}=\p(q) \mp i\uv$,
where
 $\p(q)=\sin q \left(1+\frac{u^2}{16\cos^2
q}\right)^{1/2}$. Thus  $S_{12}$ vanishes for the (singlet) bound state,
 while $S_{21}$ has a
pole. Eq (\ref{bst2}) therefore 
gives correctly the  wavefunction (\ref{bst}) after
antisymmetrization because the ``forbidden'' function $e^{\xi n}$ in
interval
$n>0$ is projected out.
One sees that a subset of S-matrices become undefined, in this case
$S_{21}$, within the Bethe ansatz for single-particle states for
$L\ra\infty$ \cite{yang2}. To avoid these singular 
S-matrices one has to treat the
bound pair as a composite with a well defined
S-matrix $S^{ub}$ with unbound states as well as with other
bound states,  $S^{bb}$. 
To compute $S^{ub}$ we write,
\be
S^{ub}_{1(23)}=S^{uu}_{13}(k,k^+)S^{uu}_{12}(k,k^-)
\label{frb}
\ee
and note
that the spin space of the three particles, $V$, is restricted: 2 and 3
form a singlet in the region
$[123]$ denoting the ordering $(n_1<n_2<n_3)$. It  
is $V=V_1\otimes
V^{singlet}_{23}$, constituting a two dimensional  
subspace of the
eight dimensional
spin space of the
three particles. In this space
$P_{23}A_{[123]}=-A_{[123]}$, so that
$P_{13}P_{12}=P_{12}P_{23}=-P_{12}$. Further
 $P_{12}=\id-P_{13}$ when acting on
$V$. In general
these operators do
not leave $V$
invariant, and 
the singlet state of
particles 2 and 3
will be destroyed
upon scattering with
particle 1. It is
due to a non trivial
cancellation of
terms for 
 momenta $k^+$,
$k^-$ satisfying
the pole condition
for $S^{uu}_{23}$,
that
$S^{ub}_{1(23)}$
indeed leaves $V$
invariant, acting as
a  phase
on the wavefunction.

We find,
\be
S^{ub}_{1(23)}(k,q)=\frac{\sin k -\p(q) -i\frac{u}{4}}
{\sin k -\p(q) +i\frac{u}{4}}
\label{sub}
\ee
In an analogous way,
\be
S^{bb}_{(12)(34)}(q_\a,q_\b)=\frac{\p(q_\a) -\p(q_\b) -i\frac{u}{2}}
{\p(q_\a) -\p(q_\b) +i\frac{u}{2}}.
\label{bbsm}
\ee
Proceeding in the same manner we find more elaborate bound states: 
bound states of  bound states. We can infer the existence of these
higher composites by looking for zeros, respectively, poles, in the
S-matrix for two two-particle bound states (\ref{bbsm}). The zero of
(\ref{bbsm}) is at 
$\p(q_\a)-\p(q_\b)=i\uz$, corresponding to complex $q$'s. 
Choosing
$\p_{1,2}^{(2)}=\p_0^{(2)}\pm i\uv$, we find the four momenta of this
``double'' bound state, the so-called quartet \cite{and}:
$\sin k_{1,2}^{(2)}=\p_{1}^{(2)}\pm i\uv$, 
$\sin k_{3,4}^{(2)}=\p_{2}^{(2)}\pm i\uv$.
The S-matrix of this state with an unbound particle is,
\be
S_{k\p_0^{(2)}}^{u(bb)}=
\frac{\sin k -\p_0^{(2)} -i\frac{u}{2}}
{\sin k -\p_0^{(2)} +i\frac{u}{2}}
\ee
In general there are bound complexes of $2m$ electrons,
($m$-complexes), which
correspond to a pole in the S-matrix of an $(m-1)$-complex
with a simple bound pair,(a 1-complex). The $m$-complex
is parameterized by an  $m$-string of the form:
\be
\p^{(m)}_{a,j}=\p_a^{(m)}+(m+1-2j)i\uv \ \ \ j=1\ldots m
\ee
The S-matrix of 
an unbound particle with a $m$-complex is
\be
S_{k\p_a^{(m)}}^{u(m)}=
\frac{\sin k -\p_a^{(m)} -im\frac{u}{4}}
{\sin k -\p_a^{(m)} +im\frac{u}{4}}
\ee
and the S-matrix of an $m$-complex
with an $n$-complex: 
\bea
 S_{\p_a^{(m)}\p_b^{(n)}}^{(m)(n)}& =&
\frac{\p_a^{(m)}-\p_b^{(n)}-|n-m|i\uv}{\p_a^{(m)}-\p_b^{(n)}+|n-m|i\uv}
\frac{\p_a^{(m)}-\p_b^{(n)}-(n+m)i\uv}{\p_a^{(m)}-\p_b^{(n)}+(n+m)i\uv}
 \label{smbs} \nn \\
 && \prod^{{\textrm{\scriptsize{min}}}(m,n)-1}_{l=1}
\left(\frac{\p_a^{(m)}-\p_b^{(n)}-(|n-m|+2l)i\uv}
{\p_a^{(m)}-\p_b^{(n)}+(|n-m|+2l)i\uv}\right)^2  
\nn
\eea

Using these S-matrices we  construct, in the usual
 manner (see e.g. \cite{and}),
 eigenstates of the hamiltonian
characterized  by: $N^u$, the number of unbound particles,
and $N^b_m, m= 1 \ldots \infty$, the number of $m$-complexes. These
are indeed eigenstates since they derive from S-matrices satisfying 
the Yang-Baxter relation.
The antisymmetrization of the wavefunction is trivial within a complex,
and  amounts to
a change of labels between the constituents of the complexes.

We now impose periodic boundary conditions and derive the 
BA equations.
Consider  a system with
 $N=N^u +2N^b$ electrons, where
$N^u$ particles are in plane wave states given by momenta $k_j$,
$j=1\ldots N^u$ and 
 $2N^b$ particles are in 
bound states, $N^b= \sum_m mN^b_m$. 
The total spin $S$ is given by
$S=\frac{1}{2}(N^u-2M^u)$,
where $M^u$ denotes the number of down spins among the $N^u$
unbound electrons. Imposing periodic boundary conditions
on the unbound particle $j$ we are led to the diagonalization of
 the matrix 
$Z_j$ which takes it around the ring, 
\be
Z_j = S^{uu}_{1j}\ldots S^{uu}_{N^uj}\prod_m S^{u(m)}_{1j}\ldots 
S^{u(m)}_{N^b_mj}
\ee
with eigenvalue $z_j=e^{ik_jL}$. Similarly, taking an $m$-complex around
the ring leads to the matrix (actually a phase) $Z_{(a,m)}$
\be
Z_{(a,m)}= \prod_j^{N^u}S_{k_j\p_a^{(m)}}^{u(m)}
 \prod_{(b,n)}S^{(n)(m)}_{\p_b^{(n)},\p_a^{(m)}}
\ee
where the index $(b,n)$ runs over the set of all complexes present (omitting
 $(a,m)$).
The 
corresponding eigenvalue is
$z_{(a,m)}=\exp{(iL[-2\Re\arcsin(\p_a+mi\uv)+(m+1)\pi])}$.
The modified boundary condition for the bound complexes
allows for a total number of states, $4^L$. This can be seen as follows:
The number of states in the system does not depend  on $u$,
but since in the limit $u\ra \infty$ the localization length of all
bound states tends to zero, see (\ref{bst}),  periodicity in the
center-of-mass coordinate of the composite becomes equivalent to
strict periodicity. From continuity follows then that the total
number of states is $4^L$ for any finite $u$. 

Diagonalizing the transfer matrices leads to the following set of equations,
\bea
e^{ik_jL}=\prod_{\d=1}^{M^u}\frac{\l_\d-\sin k_j-i\frac{u}{4}}
{\l_\d-\sin k_j+i\frac{u}{4}}
\prod_{(a,m)}S_{\p_a^{(m)},k_j}^{(m)u}
\label{rekg} \\
\prod_{\d\neq\g}^{M^u}\frac{\l_\g-\l_\d-i\frac{u}{2}}
{\l_\g-\l_\d+i\frac{u}{2}}
=\prod_{j=1}^{N^u}\frac{\l_\g-\sin k_j-i\frac{u}{4}}
{\l_\g-\sin k_j+i\frac{u}{4}}
\label{sping}\\
e^{iq^{(m)}({\p^{(m)}_a})L}=\prod_{(b,n)}S^{(n)(m)}_{\p_b^{(n)},\p_a^{(m)}}
\prod_j^{N^u}S_{k_j\p_a^{(m)}}^{u(m)}
\label{bskg}
\eea
 The equations (\ref{rekg},\ref{sping},\ref{bskg}), to which we refer to as 
Bethe Ansatz equations for Composites (BAC), determine the eigenstates
of 
the Hubbard
hamiltonian in the presence of bound states. They are {\it weaker}
than 
the "String 
Lieb-Wu equations".
 The latter include additional constraints which  lead to 
 over-determination; a quick way of establishing it
is to carry out a counting of the states
 described by the BAC equations. This can be done
along the lines of
\cite{taka,kor}, and
one finds exactly $4^L$ states, proving a forteriori
that the stronger set 
(\ref{lw1rn}-\ref{lw1arn}) is over-determined.\footnote{Takahashi \cite{taka}
obtained the set (\ref{rekg}-\ref{bskg}) from
the set (\ref{lw1rn}-\ref{lw1arn}) neglecting
the terms of order $\vep$ and eliminating the $\varphi_l$'s
by plugging (7) into (8). He 
assumed
the same number of solutions as follows from BAC. In this way he
was able to derive the correct thermodynamics.}

The BAC equations modify the structure of the upper Hubbard band. We shall
show now that in addition
to the holon-antiholon excitation which have linear dispersion at low energies
  a new excitation appears with a nonlinear dispersion. 
Consider a
single bound pair above the ground state of the repulsive Hubbard
model. To keep the number of particles  $N$ unchanged 
 two holes are placed in the sea of unbound particles. Thus,
$N^u=N-2$ and $M^u=N/2-1$ in the notation above. Equations
(\ref{rekg},\ref{sping},\ref{bskg}) become,
\bea
&&Lk_j  = \sum_{\d}^{M^u}\th_1(\sin k_j-\l_\d)
+\th_1(\sin k_j -\p(q)) +2\pi n_j  \nn \\
&&\sum_{j=1}^{N-2}\th_1(\sin k_j-\l_\g)=\sum_{\d=1}^{M^u}\th_2(\l_\d-\l_\g)
+2\pi I_\g  \nn \\
&&2qL=\sum_{j=1}^{N-2}\th_1(\p(q)-\sin k_j) +2\pi J  \nn
\eea
where $\th_n(x)= -2\tan^{-1}(\frac{x}{n}\vu).$
The range for the quantum numbers $n_j$ is $-N/2\le n_j\le N/2-1$ for
$M^u$ even, and $-(N-1)/2\le n_j\le (N-1)/2$ for $M^u$  odd.
The $I_\g$ range between $-((N-2)-M^u-1)/2$ and $+((N-2)-M^u-1)/2$.
The $n_j$ sequence contains two unfilled slots, holes,
 as the actual number of unbound
$k$-momenta is $N-2$. We denote the hole positions by $k_1^h$ and $k_2^h$.
The $I_\g$ sequence does not contain holes in
the
absence of spin excitations.
 $J$, the quantum number associated with
the bound state, is an  integer if $L-N$ is even, and a half-odd
integer if it is odd. We assume $L,N$ even in the following.
To find the limiting values for J, we consider the boundaries of the
allowed range for $q$: $\pi/2\le q\le\pi$, (resp. $-\pi\le
q\le-\pi/2$). We may treat the range for $q$ as simply connected by
shifting $-\pi/2$ to $3\pi/2$. We find, $L-\frac{N}{2}\ge J\ge \frac{N}{2}$
so that the associated mode becomes a full fledged excitation, 
independent of the two holons.
At half-filling: $(N/2)\le J\le (N/2)$, which leads to maximal restriction
of the phase space for the bound state.
To compute the spectrum  we proceed in the usual manner \cite{coll,and} 
converting the BA equations to integral equations, for notation see \cite{and}.
 One needs to calculate $\rho(k)$
the density of the $k_j$ momenta, which consists of several terms,
$\rho(k)=\rho_0(k)+L^{-1}[\rho_1(k;k^h_1)+\rho_1(k;k^h_2)+\rho^b(k;q)]$.

The first
term describes the $k_j$ distribution in the ground state, 
$\rho_1(k;k^h_j), ~j=1,2$
are the holon contributions, and 
  $\rho^b(k;q)$ is the bound-state contribution. We write here
only the integral equation for $\rho^b(k;q)$,
\bea
\rho^b(k)-\cos k \frac{4}{u}\int_{-Q}^Q \rd k'R\left(\frac{4}{u}
(\sin k-\sin k')\right)\rho^b(k') \nn \\
=\cos k~K_1(\sin k -\p(q)). \nn
\eea
Here
 $Q$ is the $k$  Fermi-level determined by the requirement
$\int_{-Q}^{Q} \rho(k) ~dk =N^u/L$.

We find that the total excitation energy and momentum consist of 
contributions from
 the two holons
and from the bound state,
\bea
\Delta E &=& E-E_0=\ep^h( k_1^h)+\ep^h( k_2^h)+\ep^{b}(q) \nn \\
\Delta P &=& P-P_0 = p^h(k_1^h)+p^h(k_2^h)+p^{b}(q) 
\eea
with direct, and a backflow  contribution. The holon
excitation is well known \cite{woyn,and}. 
In particular, $\ep^h(k^h)$
vanishes linearly as $k^h \ra Q$. This corresponds
to the holon momentum, $p^h_j=\int_0^{k_j}\rd k\r_0(k)$, tending to
the
charge Fermi-momentum $\pm\k^c_{F}=\pm 2k_F=\pm\pi(N/L)$.
The bound state energy is computed from,
\bea
\ep^{b}(q) = \E(q) -2t\int_{-Q}^Q\rd k \r^{b}(k;q)\cos k
-\mu \int_{-Q}^Q\rd k \r^{b}(k;q) \nn
\eea
with $\E(q)=-4t\cos q\cosh(\xi(q))$, and  $\mu=dE/dN$, being 
the chemical potential
\cite{coll,and}.
At half-filling $\ep^{b}(q)=U$  and the dispersion
of the total excitation depends only on the holon parameters $k_1^h,k_2^h$.
 However, away from half-filling the bound state becomes an independent
excitation with momentum,
 $
p^{b}(q)=2q-\int_{-Q}^Q\rd k\r_0(k)\th_1(\p(q)-\sin k).$
As $q$ runs over the allowed range 
$\pi/2<q<\pi$ (resp. $-\pi<q<-\pi/2$) , $p^{b}$ varies in the range $(\pi+k_F, 
2\pi)$
(resp. $(-2\pi,-\pi-k_F$). This corresponds to a
symmetric band around zero between $-|\pi-k_F|$ and $\pi-k_F$.
It follows that at half-filling $p^{b}\equiv 0$,  the
phase volume vanishes. Starting directly from
this case and setting $J=0$, $q$ becomes fixed in
terms
of the hole parameters $k^h_1$ and $k^h_2$:
 $\p(q)=\frac{1}{2}(\sin k^h_1+\sin k^h_2)$ \cite{woyn}.
This statement, however, is empty as 
the {\it physical} parameter of the excitation, the  energy and momentum, 
 are  independent of
the
unphysical parameter $q$:  $E^{b}\equiv U$ and $p^{b}\equiv
0$.   
Away from half filling the dispersion of the new mode 
 is given parametrically by $\ep^{b}(q)$ and
$p^{b}(q)$. The solution of the equation for $\rho^b$
 has to be performed
numerically in general.
However, at the band edges, $p^{b}=\pm(\pi-k_F)$, 
analytic results are accessible.
 In particular,  the velocity,
$v^b=(\rd\ep^b/\rd p^b)|_{p^{b}=\pm(\pi-k_F)}=0$.
It follows that this  excitation is quite different from the
gapless
particle-hole excitations, as the nonzero curvature of the dispersion
at the band edges
sets an intrinsic scale. Because these excitations do not appear in the
low energy regime they do not modify the transition to the
known continuum limit of the model (the 
Luttinger liquid \cite{haldane,schulz,voit}) but the parameters
of the model need to be further studied.
\cite{preparation}.

 Similar consideration lead to a modification of the (gapped) spin excitation
sector in the attractive case. Again, a new dispersive spin singlet
 mode makes its appearance in addition to the  spinons \cite{and}. 
Also
other models where complex momenta  appear, such as the 
multichannel Kondo, the Anderson or the $t-J$ model, 
need to be further examined \cite{preparation}.

Acknowledgments: We wish to thank C. Bolech, 
A. Jerez, G. Kotliar, M. Milenkovic,
 A. Ruckenstein for stimulating and enlightening discussions and
comments.\\
This work is supported in part (D.B.) by the Deutsche Forschungsgemeinschaft.

\end{multicols}

\end{document}